\documentclass[useAMS,usenatbib]{mn2e}
\usepackage{epsfig}
\usepackage{longtable}
\usepackage{times} 
\newcommand{\sw}{$Swift$}

\def \sw {{\em Swift}}

\def \apj {ApJ}
\def \apjs {ApJS}\def \aap {A\&A}
\def \mnras {MNRAS}
\def \arcmin {'}

\title[   ]
{
Timing and spectral study of the Be XRB IGR~J11305--6256: Swift discovers the orbital 
period and a soft X-ray excess.
}

\author[V.\ La Parola et al.]{V.\ La Parola$^{1}$, A.\ D'A\`i$^{2}$, G.\ Cusumano$^{1}$,
A.\ Segreto $^{1}$,  N.\ Masetti $^{3}$, A.\ Melandri$^{4}$\\
$^{1}$INAF - Istituto di Astrofisica Spaziale e Fisica Cosmica, Via U.\ La Malfa 153, I-90146 Palermo, Italy\\
$^{2}$Dipartimento di Fisica e Chimica, Universit\`a di Palermo, via Archirafi 36, 90123, Palermo, Italy\\
$^{3}$INAF - Istituto di Astrofisica Spaziale e Fisica Cosmica di Bologna, via Gobetti 101, 40129, Bologna, Italy\\
$^{4}$ INAF - Brera Astronomical Observatory, via Bianchi 46, 23807, Merate (LC), Italy \\
}

\begin{document}

\date{}

\pagerange{\pageref{firstpage}--\pageref{lastpage}} \pubyear{2013}

\maketitle

\label{firstpage}

\begin{abstract}
IGR~J11305-6256 is one of the numerous sources discovered through the INTEGRAL scan of the
Galactic Plane. Thanks to the Swift-BAT survey, that allows the frequent sampling of any sky region, 
we have discovered in the hard X-ray emission of this source a modulation with a 
period of $120.83$ d. The significance of this periodic modulation is $\sim$ 4 standard 
deviations in Gaussian statistics. We interpret it as the orbital period of the binary system.
We derive an orbital separation between IGR~J11305-6256 and its companion star of 
$\sim 286 R_{\odot}$ corresponding to $\sim 19$ times the radius of the companion star.
The broadband XRT-BAT (0.3$-$150 keV) spectrum is described either by the sum of a
black-body and a cut-off power-law or by a partially absorbed 
cut-off power-law. The temporal and spectral characteristics of the source
indicate its possible association with the class of persistent, but faint, Be X-ray  binary systems.

\end{abstract}

\begin{keywords}
X-rays: binaries -- X-rays: individual: IGR~J11305-6256. %,  IGR~J16493$-$4348, IGR~J16418$-$4532.

\noindent
Facility: {\it Swift}

\end{keywords}

%%%%%%%%%%%%%%%%%%%%%%%%%%%%%%%%%%%%%%%%%%%%%%%%

	%%%%%%%%%%%%%%%%%%%%%%%%%%%%%%%%%%%%%%%%%%%%%%%%%%%%%%%%%
	\section{Introduction\label{intro}}
	%%%%%%%%%%%%%%%%%%%%%%%%%%%%%%%%%%%%%%%%%%%%%%%%%%%%%%%%%

During the last decade, the IBIS/ISGRI telescope \citep{ubertini03}
on board the INTEGRAL gamma-ray satellite \citep{winkler03}
has discovered several new Galactic sources, thanks to the 
hard X-ray window of the telescope (20$-$80 keV) that allows to unveil the presence of 
highly absorbed objects that appear too faint in the soft X-ray energies, and thanks to the
wide field of view ($\sim30^o$) and to the scan monitoring program of the Galactic plan
that favored the capture of transient episodes from  previously unknown objects. 
Moreover, since November 2004 the Swift satellite \citep{swift}, with its Burst Alert Telescope 
(BAT, \citealp{bat}) has been performing a continuous monitoring of the hard X-ray sky 
with different observational features with respect to ISGRI: the observation 
of large sky areas thanks to a field of view of 1.4 steradian (half coded) and
several switches of the satellite pointing direction within a day. 
These characteristics allow to observe up to $\sim80\%$ of the entire sky every day, with a duty
cycle of 10 to 20\% for each direction. This resulted of paramount importance to monitor 
the  variability of the sources and to achieve good statistics for broad band spectral analysis.
In particular for the high mass X-ray binaries (HXMBs) class, it has led to 
a sensible increase of orbital period detections (e.g. \citealp{corbet1, corbet2, corbet3, corbet4, 
corbet5, corbet6, corbet7, cusumano10, laparola10, dai11a}), and to discover
absorption features in their hard X-ray spectra \citep{dai11b}.

In this Letter we analyze the soft and hard X-ray data collected by Swift on
IGR~J11305-6256.

IGR~J11305-6256 is a transient source discovered by the IBIS/ISGRI telescope 
in the Carina region, with a flux of 8 mCrab in the 20--60 keV band \citep{atel278}.
A Swift-XRT observation allowed the accurate localization of IGR J11305-6256
(RA$_{J2000}$ = 11h 31m 06.5s, Dec$_{J2000} = -62^{\circ} 56\arcmin 46\arcsec.6$,
error radius: 6\arcsec) and confirmed the association with the blue giant star HD 100199
(of spectral type B0 IIIe, \citealp{garrison77}), located at a distance of $\sim$ 3 kpc 
\citep{masetti06}.
The X-ray source position was further refined through a Chandra observation
(RA$_{J2000}$ = 11h 31m 06.95s, Dec$_{J2000} = -62^{\circ} 56\arcmin 48\arcsec.9$, with position
uncertainty $\simeq 0.1\arcsec$, \citealp{tomsick08}). The Chandra data 
showed a weakly absorbed ($N_{\mathrm{H}\,}=3.2^{+2.8}_{-2.2}\times 10^{21}\
\mathrm{cm}\,^{-2}$) flat  power law ($\Gamma=0.33^{+0.40}_{-0.28} $) spectrum, with 
a 0.3-10 keV unabsorbed flux of $44^{+20}_{-34} \times10^{-12}$ erg cm$^{-2}$ s$^{-1}$,
corresponding to a luminosity of $4.7^{+2.1}_{-3.6}\times 10^{34}$ erg s$^{-1}$, at a
distance of 3 kpc.
This Letter is organised as follows. Section 2 describes the Swift data 
reduction. Section 3 reports on the timing analysis. 
Section 4 describes the broad band spectral analysis.  In Section 5 we briefly 
discuss our results.

	%%%%%%%%%%%%%%%%%%%%%%%%%%%%%%%%%%%%%%%%%%%%%%%%%%%%%%%%%
	\section{Data reduction\label{data}}
	%%%%%%%%%%%%%%%%%%%%%%%%%%%%%%%%%%%%%%%%%%%%%%%%%%%%%%%%%
\begin{figure*}%%%%%%%%%%%%%%%%%%%%%%%%%%%%%%%%%%%%%%%%%%%%%%%%%%%%% 
\begin{center}
%\vspace{-1.5truecm}
\centerline{\includegraphics[width=14.cm,angle=0]{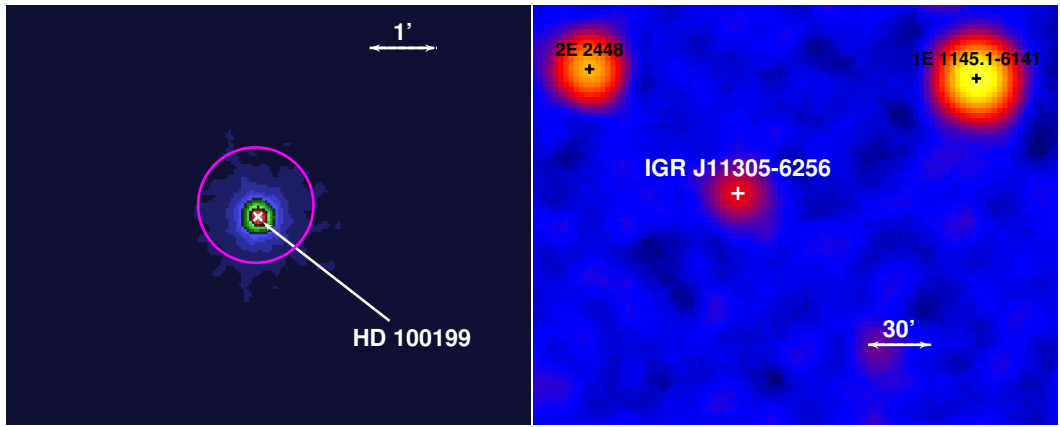}}
\caption[IGR~J11305-6256 sky maps]{ 
Left panel: 0.2$-$10 keV XRT image with superimposed the position of the optical
counterpart, marked with a cross, and the BAT error circle of 0.86'
(magenta circle). Right panel: 15$-$150 keV BAT significance 
map in the sky region around IGR~J11305-6256.}
\label{map} 
\end{center}
\end{figure*}

\begin{figure}
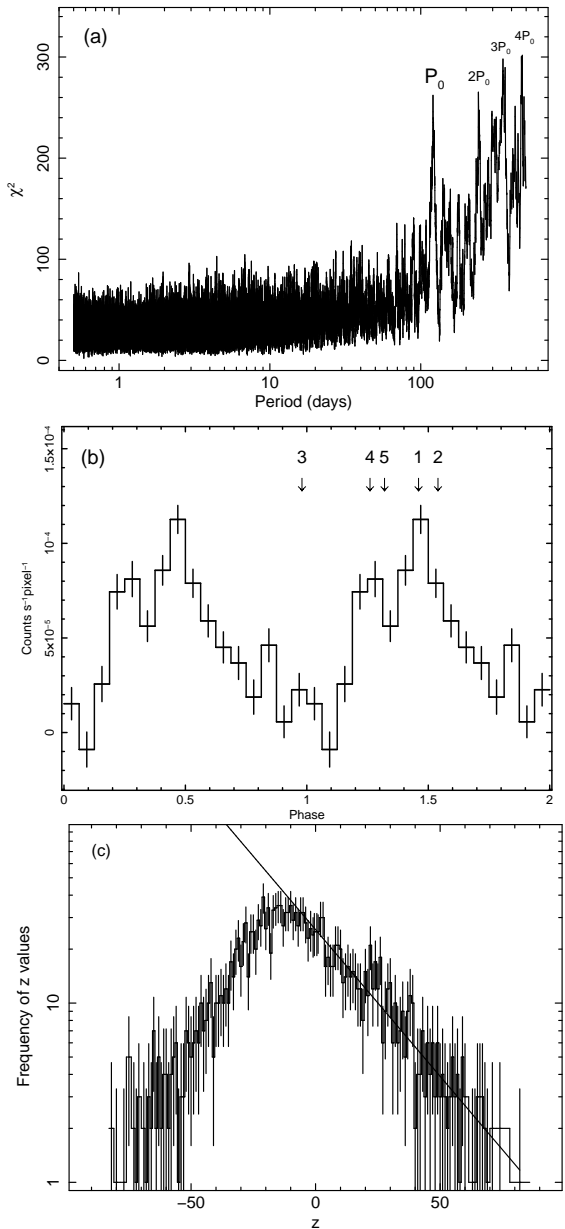
%%%%%%%%%%%%%%%%%%%%%%%%%%%%%%%%%%%%%%%%%%%%%%%%%%%%% PAP VII FIGURE 2
\begin{center}
%\vspace{-1.5truecm}
\centerline{\includegraphics[width=5.4cm,angle=270]{figura2.ps}}
\centerline{\includegraphics[width=5.4cm,angle=270]{figura4_new.ps}}
\centerline{\includegraphics[width=5.4cm,angle=270]{figura3.ps}}

%\vspace{-2.5truecm}
\caption[]{{\bf a}: Periodogram of \sw-BAT (15$-$150\,keV) data for 
IGR~J11305-6256.
{\bf b}: Light curve folded at a period $P= 120.83$\,day, with 16 phase 
bins. The vertical arrows mark the phases corresponding to the XRT observations.
{\bf c}: Distribution of $z=\chi^2-F_{\chi}$ values extracted in the period range
between 20 and 200 days, excluding the z values around  $P_0$.
The continuous line is the best fit obtained with an exponential model applied to the tail of the distribution above $z >20$.}               
                \label{period} 
        \end{center}
        \end{figure}

Swift-XRT \citep{xrt} observed IGR~J11305-6256 five times between November and December 2005. 
The source was always observed in photon counting mode \citep{hill04}.
Table \ref{log} reports the log of these observations with the most relevant details.
XRT data were calibrated, filtered and screened using the standard procedures 
included in the XRTDAS package HEAsoft 6.0.4.
For each observation we extracted source events and spectra from a circular region of 20 pixel radius (1
pixel = 2.36'') centered on the source position as determined with
{\sc xrtcentroid}, adopting standard grade filtering 0$-$12.
Fig.\ref{map} (left) shows the 0.2$-$10 keV XRT image.
The event arrival times were converted to the solar system barycentre (SSB) time with the task {\sc barycorr}.
The background spectra were built using the same grade selection and selecting events  in an annular region
centered on the source with radii of 40 and 70 pixels to avoid contamination from the 
source point spread function wings.
The source spectra of each observation were summed to obtain a
single spectrum, and the same was done for the background spectra. The 
ancillary files were combined using {\sc addarf}
weighting them by the exposure times of the relevant spectra. Finally, spectra
were re-binned with a minimum of 20 counts per energy channel to allow the use of
the $\chi^2$ statistics. The spectra were analyzed  using the  spectral 
redistribution matrices and the ancillary response file v.011 (suitable for 
data collected with CCD substrate voltage $\rm V_{ss}=0V$).

\begin{table}
%\scriptsize
\begin{tabular}{l l l r r r r}
\hline
Obs & Obs ID     & $T_{start}$  & $T_{elapsed}$  & $T_{exp}$ & $Ph_{Orb}$ & rate\\
 &            &  (MJD)               & (s)            &  (s) & & c/s\\  \hline
1&35098001 & 53627.01	  & 81703    & 6007 & 0.46 &  0.33\\
2&35224002 & 53636.33	  & 12118    & 1048 & 0.54 &  0.38\\
3&35224003 & 53690.03	  & 82708    & 15899& 0.98 &  0.43\\
4&35098002 & 53725.67	  & 23856    & 3658 & 0.28 &  0.14\\
5&35098003 & 53728.00	  & 19014    & 5127 & 0.30 &  0.13\\
\hline
% 4 source 1  back     1.3 sigma    8.88275
\end{tabular}
\caption{Log of Swift-XRT observations. $T_{elapsed}$ is the observation lenght; $T_{exp}$ is
the net exposure time; $Ph_{Orb}$ is the orbital phase referred to the profile in
Figure~\ref{period}b.}
\label{log}
\end{table}

The raw BAT survey data of the first 88 months of the Swift
mission were retrieved from the HEASARC public
archive\footnote{http://heasarc.gsfc.nasa.gov/docs/archive.html} and
processed with the {\sc batimager} software \citep{segreto10}, that performs 
screening and mosaicking of the survey data and produces  
 background subtracted spectra and light curves for each detected source.
Fig.\ref{map} (right) shows the 15$-$150 keV significance sky map (exposure time
of 17.7 Ms) centered on IGR~J11305-6256, where the source is detected with a maximum significance
of 26.7 standard deviations. The light curve for timing analysis
was extracted in the same energy range.
The time tag of each bin was corrected to the solar system barycentre (SSB) by using the task
{\sc earth2sun}. The official BAT spectral redistribution
matrix\footnote{http://heasarc.gsfc.nasa.gov/docs/heasarc/caldb/data/swift/bat/index.html}
was used for spectral analysis. Quoted errors are given at 90\% confidence level for 
a single parameter, unless otherwise stated.

	%%%%%%%%%%%%%%%%%%%%%%%%%%%%%%%%%%%%%%%%%%%%%%%%%%%%%%%%%

	\section{Timing analysis\label{timing}}
	%%%%%%%%%%%%%%%%%%%%%%%%%%%%%%%%%%%%%%%%%%%%%%%%%%%%%%%%%

In order to search for long periodicities in the emission of IGR~J11305-6256,
we exploited the 15$-$150 keV BAT light curve using the folding technique \citep{buccheri85}:
we searched in the 0.5$-$500 days period range, with a period resolution of  P$^{2}/(N 
\,\Delta$T$_{\rm BAT})$, where P is the trial period, $N=16$ is the number of phase bins used to
build the profile and $\Delta$T$_{\rm BAT}\sim 231$ Ms is the  data time span. 
The average count rate in each profile bin has been evaluated weighting the rates by the inverse square of their
statistical error, which is appropriate when dealing with a large spread in the error values.
This happens in the BAT data because the source is typically observed at several off-axis
angles, thus introducing a large spread in the source signal-to-noise ratio (SNR), uncorrelated with the source count rate.

Fig.~\ref{period}~(a) shows the periodogram where a strong feature at
P$_0=120.83\pm0.34$ d (the error is the period resolution at P$_0$) emerges with $\chi^2\sim
261.8$. Other relevant peaks are visible for periods corresponding to multiples of
P$_0$. The pulsed profile evaluated at P$_0$ with T$_{\rm epoch}=
54658.949 $ MJD is shown in Figure~\ref{period}~(b). 
The periodogram is characterised by an increasing strong red noise for higher period values, caused by the
long term variability of the source. For this reason the significance of the
periodicity at P$_0$ cannot be evaluated relying on the $\chi^2$ statistics, and an alternative method
shall be used. We proceeded according to the following steps.

\begin{itemize}
\item We modeled the ascending trend of the $\chi^2$ distribution fitting it with a 2nd order
polynomial and we created a new distribution (z) subtracting the best fit $F_{\chi}$ from the 
$\chi^2$ distribution. The z value at $P_0$ is $\sim 173.4$. 
\item We built the histogram of the z distribution (Figure~\ref{period}, c) 
extracting the z values only from  the period range 60--200~days and excluding the interval around $P_0$.
\item We fitted the histogram values for $z >20$  with an exponential 
function. The resulting best fit model is shown in  Fig.~\ref{period} (c). 
\item We evaluated the area under the histogram: we  summed the area of each 
single bin from its left boundary up to $z=20$; beyond $z=20$  we integrated the best fit 
exponential model up to infinity. 
\item We evaluated the integral of the best-fit exponential 
function beyond 173.4 and normalised it to the total area of the histogram. 
\end{itemize}

The result ($1.2\times10^{-4}$) is the probability of chance occurrence of a 
z value equal to or larger than 405 (or a $\chi^2$ equal to or larger than 261.8) and it 
corresponds to a significance for the  detected feature of $\sim 3.9$ standard deviations in Gaussian statistics. 

\begin{figure}%%%%%%%%%%%%%%%%%%%%%%%%%%%%%%%%%%%%%%%%%%%%%%%%%%%%% PAP VII FIGURE 2
\begin{center}
\centerline{\includegraphics[width=5.4cm,angle=270]{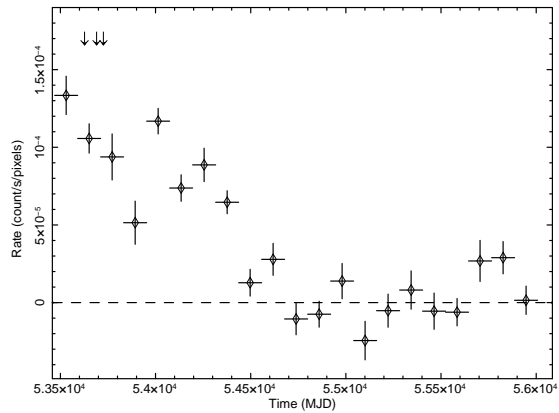}}
\caption[]
{88-month BAT light curve. The bin length corresponds to the period 
$P_0= 120.83$\,day. The vertical arrows mark the epoch of the XRT observations. 
The first and the third bars correspond to XRT observations 1-2 and 4-5 respectively.}               
                \label{lc} 
        \end{center}
        \end{figure}

Fig.~\ref{lc} shows the BAT 15$-$150 keV light curve sampled at time intervals equal to P$_0$. 
The source shows a strong long term variability. Significant X-ray emission was observed up to mid 2007
($\sim$ MJD 54400), then the source weakened to an intensity level averaged over P$_0$
roughly consistent with the background.

IGR~J11305-6256 has been always detected in the XRT pointed observations. Table~\ref{log} reports the source
count rate and the orbital phase of each observation evaluated with respect to P$_0$ and 
T$_{\rm epoch}$. The source intensity varies between 0.1 and 0.4 counts/s and these variations 
result uncorrelated to the P$_0$ modulation shown in Fig.~\ref{period}~(b).
The timing analysis on the XRT data has to cope with two main issues: the read-out time of
the XRT-PC data ($\delta T_{XRT}$=2.5073 s) and the fragmentation of the observation into several
snapshots with different duration and time separation. These effects introduce systematic features in
the results of the timing analysis that can mask the significance of any real source modulation.
Therefore we performed a folding analysis for each XRT observation on the events
arrival times randomized within the XRT-PC time resolution bin. 
Moreover, we produced the periodogram relevant to each snapshot with exposure time higher than
500 s searching in the period range $\delta
T_{XRT}-100$ s: the periodograms obtained from snapshots belonging to the same observation were finally summed. 
We did not find any significant feature in the resulting five periodograms.
We repeated this analysis  selecting snapshot lasting more than 1000 s and 
searching for modulations up to 250 s. Again, no significant features emerged above the white noise.

	%%%%%%%%%%%%%%%%%%%%%%%%%%%%%%%%%%%%%%%%%%%%%%%%%%%%%%%%%

	\section{Spectral  analysis\label{spectral}}
	%%%%%%%%%%%%%%%%%%%%%%%%%%%%%%%%%%%%%%%%%%%%%%%%%%%%%%%%%
	
In order to have the best available SNR
spectrum, we merged all the XRT spectra from the observations listed in Table~\ref{log}
and  we accumulated a hard X-ray BAT spectrum over the 88-month of monitoring.  
We preliminarily checked  that the variability of the spectral parameters 
was lower than the constraints derived by a fit error determination on the single
observations of Table\ref{log}.
To this aim, the background subtracted spectra of the XRT observations were fitted simultaneously with a model
consisting of an absorbed power-law. Forcing the spectral 
parameters to common values for the five datasets and allowing only a 
multiplicative constant to weight for 
different fluxes, we obtained similar residuals.

% forcing the photon index and the absorption column density to have the same value 
% for the five datasets, thus allowing only the normalization to vary for each spectrum. 
%The residuals showed a significant excess below 1 keV. 
%We tried several models to account for this excess finding that it can be modeled with a 
%partial covering absorption, with $\rm n_H=0.90_{-0.11}^{+0.12}$ and covering fraction 
%$f_{nH}=0.909_{-0.016}^{+0.015}$, over a power-law with photon
%index $\Gamma=0.98\pm0.06$. The residuals show the same trend for all the data.

Several BAT spectra were produced selecting the data in different time intervals (MJD intervals 
53470$-$53832, 53832$-$54436, and 54436$-$56006; see Fig.~\ref{lc}) and 
on three different orbit phase intervals (0.19$-$0.56, 0.56$-$0.87, and 0.87$-$1.19, see  Fig.~\ref{period}-b). 
These spectra were fitted  with a power-law model and, as above, forcing a common photon index
for all the spectra. The best fit residuals show the same trend for all the datasets, with best fit photon index $2.0\pm0.2$.

\begin{figure}%%%%%%%%%%%%%%%%%%%%%%%%%%%%%%%%%%%%%%%%%%%%%%%%%%%%% PAP VII FIGURE 2
\begin{center}
\centerline{\includegraphics[width=8cm,angle=0]{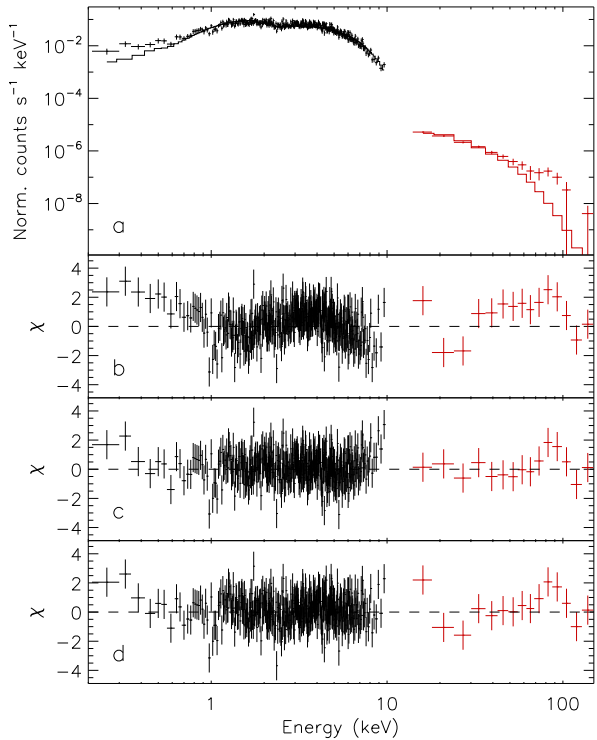}}
\caption[]
{IGR~J11305-6256 broad band spectrum. {\bf Panel (a)}: 
XRT and BAT data, best-fitting {\tt phabs*cutoffpl)} model (red line). {\bf Panel (b)}
Residuals in units of $\sigma$ for {\tt phabs*cutoffpl)} model.
{\bf Panel (c)}: Residuals in unit of $\sigma$ for the {\tt  phabs*(bbody+cutoffpl)} model.
{\bf Panel (d)}: Residuals in unit of $\sigma$ for the {\tt phabs*pcfabs*(cutoffpl)} model.
               }               
                \label{spec} 
        \end{center}
        \end{figure}

The broad band average spectrum of IGR~J11305-6256 (XRT: 0.3$-$10 keV; BAT: 15$-$150 keV)
is plotted in Fig.~\ref{spec} (a). We introduced a multiplicative factor in the model 
to account for the non-simultaneity of the BAT and XRT spectra and for any inter-calibration uncertainty. 
The factor is kept fixed to 1 for the XRT data and left free to vary for the BAT data.
First we tried an absorbed power-law with a high-energy cut-off (model 1). 
The reduced $\chi^2$ is 504.4 (with 389 dof) 
with residuals below 1 keV strongly suggesting the presence 
of other continuum components. Therefore, we added to the power-law emission 
a black-body component at a temperature of $\sim$ 1.6 keV (model 2), obtaining a 
significant improvement of the fit residuals (reduced $\chi^2$ = 371.5 for 387 dof).
In order to evaluate the statistical significance of this improvement 
we built 100000 simulated XRT spectra using the best fit parameters of model 1 
and the rate uncertainties of the observed XRT data. The simulated spectra were fitted
together with the BAT spectrum both with model 1 and model 2 obtaining a 
value of $\sim65$ as the highest difference in the $\chi^2$ values for the two models.
This results corresponds to a chance probability to obtain the decrement in $\chi^2$ as measured 
in the observed XRT+BAT source data lower than $1.0 \times 10^{-5}$, that corresponds to a
significance larger than $\sim 4.4$ standard deviations in Gaussian statistics.
In the 0.5$-$100 keV energy range 
the fluxes of the black-body component and of the power-law emission are 
3.4$\times$10$^{-11}$ and 4.0$\times$10$^{-11}$ erg cm$^{-2}$ s$^{-1}$, respectively 
(assumed with respect to the XRT data).
Alternatively, we also explored a scenario where the continuum X-ray emission (modeled with only 
a cut-off power-law) is partially absorbed  by local neutral matter (\texttt{pcfabs} in
Xspec; model 3).
This model gave a statistically  similar result  (reduced $\chi^2$ = 0.97 for 386 dof), compared
to the {\tt blackbody + cutoff powerlaw}. We present in Table \ref{fit}, the 
best-fitting parameters for the spectral models that have been considered, while
data and residuals in units of sigmas are shown in Fig.\ref{spec}.

\begin{table*}[htp]
\caption{Best-fitting spectral parameters. $\dag$ Covering fraction for N${_\textrm{H}}$ 2. $\ddag$
Constant factor to be multiplied to the model in the BAT energy range in order to match
the BAT count rate.}
\label{fit}
\begin{tabular}{ r lll}
\hline
Parameter &   \textit{cutoff} & \textit{bb+cutoff} & \textit{pcfabs*cutoff}     \\ 
          &      (model 1)      &   (model 2)          &  (model 3)      \\ \hline \hline

N$_{\textrm{H}}$ 1  ($\times 10^{22} $ cm$^{-2}$) & 0.32 $\pm$0.05 & 0.13$\pm$ 0.08 & 0.26$\pm$0.06 \\
%pcfabs
N${_\textrm{H}}$ 2  ($\times 10^{22} $ cm$^{-2}$)    &  & &  2.6$\pm$0.6 \\
$\rm F_{n_H 2}^{\dag}$                               &  & & $0.74^{+0.04}_{-0.05}$  \\ 
%cutoffpl
$\Gamma$                          & 0.43$\pm$0.08  &  1.0$\pm$0.3       & 1.20$\pm$0.17 \\
$\rm E_{cutoff}$ (keV)            &  12.0$\pm$1.7  &  23$_{-7}^{+10}$       & 23$\pm$5 \\
$\rm N_{power law}$ ($\times 10^{-3}$ ph keV$^{-1}$cm$^{-2}$s$^{-1}$ at 1 keV)  & 1.5$\pm$0.1 & 0.4$\pm$0.2 & $ (4.7^{+1.4}_{-1.0})$ \\ 
%bbody
kT$_{BB}$ (keV)   & & 1.60$\pm$0.06  & \\
BBody radius (km) & & 0.19$\pm$0.02  & \\
%costanti varie
$\rm C_{BAT}^{\ddag}$        & 0.49$\pm$0.08  & 3.4$_{-1.2}^{+1.5}$ & $0.70^{+0.12}_{-0.11}$ \\ 
%$\rm F_{0.2-10}^{\ddag}$     &   &  & $4.7^{+0.5}_{-0.7}$ erg cm$^{-2}$ s$^{-1}$\\ 
%$\rm F_{15-150}^{\ddag}$     &   &  & $3.0^{+0.2}_{-0.3}$ erg cm$^{-2}$ s$^{-1}$ \\ 
$\chi^2$  / dof                    & 503 / 388   &  369 / 386  & 374 / 386 \\  \hline \hline
\end{tabular}
\end{table*}

%%%%%%%%%%%%%%%%%%%%%%%%%%%%%%%%%%%%%%%%%%%%%%%%%%%%%%%%%%
\section{Discussion\label{discuss}}
%%%%%%%%%%%%%%%%%%%%%%%%%%%%%%%%%%%%%%%%%%%%%%%%%%%%%%%%% 

We have presented in this work the spectral and timing results from the 
complete set of Swift observations of IGR~J11305-6256. The source is 
an accreting Be XRB, whose X-ray activity has turned to quiescence in recent years. 
Analyzing the long-term BAT light curve, we found evidence at $\sim$ 4 $\sigma$ for 
the orbital period of the system to be 120.8 days. 
The third Kepler's law allows us to derive the semi-major
axis of this binary system: 
\begin{equation}
a=(G P_0^2~(M_{\star}+M_{\rm X})/4\pi^2)^{1/3} \simeq 286 R_{\odot} = 19 R_{\star}  \\
\end{equation}
where $M_{\rm X}=1.4 M_{\odot}$ is the mass of the neutron star, 
$M_{\star} \simeq 20 M_{\odot} $ is the mass expected for the spectral type 
of the companion star  and $R_{\star} \simeq 15 R_{\odot} $ is its expected 
radius \citep{lang92}.
Such a large orbital separation is indeed common in Be XRBs.
Considering the whole class of Be XRB, the long orbital
period would suggest also very long spin periods and high eccentricity \citep{reig11}.
However, the folded light curve shows a smooth modulation with the orbital phase, and
suggests that 
accretion may not be clocked with a periastron passage, as typical for classical 
highly-eccentric Be XRBs.

The broadband data collected by pointed Swift observations and by the long-term BAT 
monitoring  indicated that the spectrum 
cannot be described with an absorbed power-law with a high-energy cut-off 
as in the case of most low-luminosity accreting high-mass/Be X-ray binaries.
The fit is sensibly improved when a thermal black-body component is added 
to the model, or when the X-ray emission is only partially absorbed.
In the first scenario, we may be observing thermal emission from the accreting
magnetic caps of the neutron star, the derived black-body radius being 
compatible with a small portion of the surface of the NS. 
The black-body 
temperature (1.6 keV) is also within the expected range for other similar Be XRB sources
for which this component was clearly detected \citep{lapalombara13}, but it is
sensibly lower than the typical temperatures for soft excesses found in other
HMXBs, where the emission may be related to the reprocessed emission at the 
magnetospheric boundary or by diffuse gas around the X-ray system \citep{hickox04}.
However, all the Be XRBs that have shown presence of this component are also 
X-ray pulsars with a high modulated fraction. Lack of evidence in the
Swift XRT data of any pulsation below 250 s may point to very long 
periods that can be detected only accumulating observations with 
very long exposures.

Another scenario, that is statistically equivalent,
explains the deviations in the soft part of the spectrum as due to the presence of
a neutral absorber that partially covers the X-ray emission. 
The covered fraction is very high ($f > 70\%$) and the relative
column density (2.6$\times$10$^{22}$ cm$^{-2}$) is sensibly higher than the whole column density due to the ISM
(0.26$\times$10$^{22}$ cm$^{-2}$). This scenario requires that the X-ray source may be
embedded (but not completely) in a dense stellar wind of the companion star, or that part of
the outer atmosphere of the companion is stripped, rapidly cooled and dispersed along the NS orbit. 
In many HMXBs this scenario has been often invoked \citep[see e.g.][]{naik11}, but rarely for 
Be XRBs, due to the fact that Be stars are less efficient in driving strong winds. 
Another important finding that would corroborate this scenario, would be the detection 
of a strong fluorescent iron line, that should be imprinted in the 
circum-stellar matter. To this aim, a high signal-to-noise at the  6$-$7 keV range observation 
would be needed. 

%%%%%%%%%%%%%%%%%%%%%%%%%%%%%%%%%%%%%%%%%%%%%%%%%%%%%%%%%
\section*{Acknowledgments}
This work has been supported
by ASI grant I/011/07/0. 
%%%%%%%%%%%%%%%%%%%%%%%%%%%%%%%%%%%%%%%%%%%%%%%%

\bsp

\label{lastpage}

\end{document}